\documentclass[12pt]{article}
\usepackage{graphicx}
\usepackage[bookmarksnumbered,colorlinks,plainpages]{hyperref}

\begin{document}

\title{Back to Maupertuis' least action principle for dissipative systems: not all motions in Nature are most energy economical}
\author{Qiuping A. Wang\\
{\small IMMM, UMR CNRS 6283, Universit\'e du Maine, }\\
{\small Anevue Olivier Messiean, 72085 Le Mans, France}}

\date{}

\maketitle

\date{}

\begin{abstract}
It is shown that an oldest form of variational calculus of mechanics, the Maupertuis least action principle, can be used as a simple and powerful approach for the formulation of the variational principle for damped motions, allowing a simple derivation of the Lagrangian mechanics for any dissipative systems and an a connection of the optimization of energy dissipation to the least action principles. On this basis, it is shown that not all motions of classical mechanics obey the rule of least energy dissipation or follow the path of least resistance, and that the least action is equivalent to least dissipation for two kinds of motions : all stationary motions with constant velocity and all motions damped by Stokes drag.
\end{abstract}


\vspace{2 cm}

The principle of least action (PLA) was for the first time clearly formulated by Maupertuis in 1744\cite{Maupertuis} with the following definition of an action $A_M=\int_a^b pdx=\int_{t_a}^{t_b} p\dot xdt$ where $p=m\dot{x}$ is the momentum, $\dot{x}$ the velocity and $m$ the mass of a body moving along $x$ axis from a point $a$ at time $t_a$ to another point $b$ at time $t_b$. He called his principle a metaphysical principle because he thought that the space integral of the momentum along the trajectory of the motion, $A_M$, represents the effort or the cost of the production of the motion\footnote{C'est cela, c'est cette quantit\'e d'action qui est ici la vraie d\'epense de la Nature, et ce qu'elle m\'enage le plus qu'il est possible dans le mouvement de la lumi\`ere \cite{Maupertuis}.} which the intelligence of the Nature must minimize. There was no explanation why $pdx$ is the effort or the cost of a motion. Thereafter, Euler applied this principle to mechanical motion\cite{Euler} with explicit use of the condition of energy conservation which Maupertuis has also implicitly used in his 1746's paper for elastic collision. Nowadays, it is a consensus that Maupertuis principle applies to energy conservative systems, i.e., the motion does not cost anything in term of energy. Clearly, the statement that $A_M$ is the cost of a motion does not hold.

Today it is well understood that the Maupertuis principle of least action (MPLA) is equivalent to the Hamilton principle of least action (HPLA), each of them having different constraints of variational calculus\cite{Gray,Gray2}. The Legendre transformation $L=p\dot{x}-H$ implies $A=A_M-T\bar{H}$  where $L$ is the Lagrangian, $H$ the Hamiltonian, $T=t_a-t_b$ the duration of the motion, $A=\int_{t_a}^{t_b} Ldt$ the action, and $\bar{H}=\frac{1}{T}\int_{t_a}^{t_b} Hdt$ the time average of $H$. Then it is straightforward to write the variation relation $\delta A+\bar{H}\delta T=\delta A_M-T\delta\bar{H}$, and to derive Newtonian equation either by the MPLA $\delta A_M=0$ under the condition of constant energy $\delta H=0$, or by the HPLA $\delta A=0$ under the condition of constant duration of motion $\delta T=0$\cite{Arnold,Stoltzner,Lanczos}.

The subsequent development of the PLA has largely favored HPLA which has been successfully applied not only to the entire classical physics but also to the quantum physics within the path integral formalism\cite{Feynman}. This success has given to HPLA a conceptual priority to all other principles, empirical laws and differential equations in different branches of physics, and inspired three major projects of its extension. The first one is to deepen the understanding of nature through this principle and to search for the fundamental meaning of its exceptional universality in physics\cite{Stoltzner,Mach,Lanczos}. The second one is to extend it to more domains such as thermodynamics, statistical mechanics (with the pioneer effort, though unfruitful, of Boltzmann, Helmholtz and Hertz\cite{Stengers}), large deviation theory\cite{Freidlin} and stochastic mechanics\cite{Yasue}. The third one is to formulate it, within classical mechanics, for damped motion of dissipative systems\cite{Gray,Sieniutycz,Vujanovic,Herrera,Galley}.

Dissipative motion has been treated for the first time by Euler \cite{Euler,Goldstine} in the formulation of the calculus of variation for Brachistochrone problem with friction, the latter being written as a nonlinear function of the square of the velocity whose physical sense is not explicit. This effort was followed by Rayleigh\cite{Goldstein}, with more physical consideration, in the proposition of a `dissipative function' $D=\frac{1}{2}\zeta \dot{x}^2$, for the special case of the Stokes' law with the drag force $f_d=-m\zeta\dot{x}$, to write $\frac{d}{dt}\left(\frac{\partial L}{\partial \dot{x}}\right)-\frac{\partial D}{\partial \dot{x}}-\frac{\partial L}{\partial x}=0$, where $\zeta$ is the drag constant and $m$ the mass of the damped body. Nevertheless, this `dissipative function' does not have any relationship with the energy dissipated by the drag force. The subsequent efforts during a long period \cite{Vujanovic,Goldstein} have led to many `dissipative Lagrangian function' \cite{Galley,Bateman,Sanjuan,Riewe,Duffin,Chandrasekar} all suffering from the shortcomings such as the non uniqueness, non universality, absence of clear physical meaning and of close energy connection like $L=K-V$ \cite{Gray,Sieniutycz,Vujanovic}.

In a recent work \cite{Wang,Wang2}, we have proposed a simple and universal Lagrangian for any dissipative force and formulated the HPLA for dissipative systems. The essential of this work is the idea of an isolated (hence Hamiltonian) total system including the damped moving body and its environment, coupled to each other by dissipative force, with a total Hamiltonian composed of the kinetic energy, the potential energy of the body, and the energy lost by the body into the environment due to dissipation. For simplicity, we suppose that the $1D$ body (system 1) is large with respect to the particles of the environment, and that it moves along the axis $x$ with velocity $\dot{x}$. Its environment (system 2 composed of $N$ particles with positions $x_i$ and velocities $\dot{x}_i$ and $i=1,2,...,N$ ) includes all the parts coupled to system 1 by friction and receiving the dissipated mechanical energy. The energy transfer from system 1 to system 2 occurs only through a friction force. The total Hamiltonian reads $H=K_1+V_1+K_2+V_2+H_{int}$ where $K_1=\frac{1}{2}m\dot{x}^2$ is the kinetic energy and $V_1$ the potential energy of the system 1, $K_2=\frac{1}{2}\sum_{i}^{N}m_i\dot{x}_i^2$ the kinetic energy and $V_2(x_1,x_2,...x_N)$ the potential energy of system 2, and $H_{int}$ the interaction energy between the system 1 and 2. $H_{int}$ is responsible for the friction law and determined by the coupling mechanism on the interface between the moving body and the environment. We can suppose that the coupling mechanism, the interface (body's shape and size, body-environment distance, nature of the closest parts of the environment to the interface, etc.) and the friction laws do not change with the virtual variation of path of the damped body. In this case, $H_{int}$ is a constant of variation and can be neglected. The Hamiltonian is then $H=K_1+K_2+V_1+V_2$ or $H=H_1+H_2$ where $H_1=K_1+V_1$ is the total energy of the system 1 and $H_2=K_2+V_2$ the total energy of the system 2 :
\begin{equation}                                            \label{e0}
H_2=\sum_{i}^{N}\frac{1}{2}m\dot{x}_i^2(t)+V_2[x_1(t),x_2(t)...x_N(t)]
\end{equation}
On the other hand, the energy of system 2 can be written as $H_2=H_2^a+E_d$ where $H_2^a$ is its energy (a constant) at $t_a$ and $E_d$ the energy dissipated from system 1 to 2 up to a time moment $t$ ($t_a\leq t \leq t_b$). $E_b$ is given by the work of the friction force $\vec{f}_d$ in the following way:
\begin{equation}                                            \label{e1}
E_d=-\int_{x_a}^{x(t)}f_d(\tau)dx(\tau)
\end{equation}
where $\tau$ is any time moment between $t_a=0$ and $t$. According to the second fundamental theorem of calculus\cite{Calculus} $F(x(t))=\frac{\partial}{\partial x(t)}\int_{x_a(0)}^{x(t)}F(\tau)dx(\tau)$, we get $f_d(t)=-\frac{\partial E_d}{\partial x(t)}=-\frac{\partial H_2}{\partial x(t)}$. Although this relation looks like the relation $f_c=-\frac{\partial V_1}{\partial x(t)}$ for the conservative force $f_c$ acting on the system 1, $E_d$ or $H_2$ cannot be considered as a potential energy of the system 1 for the following reasons: on the one hand, it depends on the history of the motion, on the other, it cannot be directly and completely converted back into kinetic energy of the system 1.

The Lagrangian $L$ of the whole system can be defined by using the Legendre transformation \cite{Wang,Wang2} :
\begin{equation}                                            \label{e1a}
L=p\dot{x}+\sum_{i}^{N}p_i\dot{x}_i-H=K_1-V_1+2K_2-H_2
\end{equation}
where $p$ is the momentum of system 1 and $p_i$ the momentum of the particle $i$ of the system 2. The corresponding action is $A=\int_{t_a}^{t_b}Ldt$ which has been used for a general formulation of the HPLA for dissipative systems \cite{Wang}. The extremum property of $A$ was verified by numerical simulation of damped motion in \cite{Wang2}.

One of the aims of the formulation of PLA for dissipative systems is to study the connection between PLA and the extremum property of energy dissipation such as the maximum or minimum dissipation, least distance or least resistance\cite{Sieniutycz,Vujanovic,Gyarmati}. However, with the formulation of HPLA mentioned above, we do not see clear hint to this connection. This is why we have thought of another form of PLA : the Maupertuis principle of least action and its formulation for dissipative motion. In what follows, we will show first of all how to make the calculus of variation with the $A_M$ defined for the whole conservative system :
\begin{equation}                                            \label{e2a}
A_M=\int_{t_a}^{t_b} [p\dot{x}+\sum_{i}^{N}p_i\dot{x}_i]dt.
\end{equation}
The second term with the summation in the integral of $A_M$ being independent of $x$, the variation of $A_M$ due to a tiny change $\delta x(t)$ of the path of the damped body is given by
\begin{equation}                                            \label{e2}
\delta A_M=\delta\int_{t_a}^{t_b} p\dot{x}dt=\int_{t_a}^{t_b}(\dot{x}\delta p+p\delta \dot{x})dt.
\end{equation}
where the first term is just $\int_{t_a}^{t_b}\delta(\frac{p^2}{2m})dt=\int_{t_a}^{t_b}\delta K_1dt$ and the second term becomes $\int_{t_a}^{t_b}p\delta \dot{x}dt=p\delta{x}|_{t_a}^{t_b}-\int_{t_a}^{t_b}\frac{dp}{dt}\delta{x}dt=-\int_{t_a}^{t_b}m\ddot{x}\delta xdt$ with the condition $\delta x(a)=\delta x(b)=0$. Eq.(\ref{e2}) now reads
\begin{equation}                                            \label{e2b}
\delta A_M=\int_{t_a}^{t_b}[\delta K_1-m\ddot{x}\delta x]dt.
\end{equation}
Now we introduce the constraint of conservation of total energy $\delta H=\delta H_1+\delta H_2=\delta K_1+\delta V_1+\delta H_2=0$ or $\delta K_1=-\delta V_1-\delta H_2=0$. Eq.(\ref{e2}) becomes $\delta A_M=\int_{t_a}^{t_b}[-\delta V_1-\delta H_2-m\ddot{x}\delta x]dt$, or
\begin{equation}                                            \label{e3}
\delta A_M=\int_{t_a}^{t_b}\left[-\frac{\partial V_1}{\partial x}-\frac{\partial H_2}{\partial x}-m\ddot{x}\right]\delta xdt
\end{equation}
which implies that the Maupertuis principle $\delta A_M=0$ necessarily leads to the Newtonian equation of damped motion:
\begin{equation}                                            \label{e4}
m\ddot{x}=-\frac{\partial (V_1+H_2)}{\partial x}=f_c+f_d.
\end{equation}
This is a general formulation of MPLA for damped motion subject to any friction force.

The stationarity $\delta A_M=0$ must be a minimum since the integral of $A_M$ in Eq.(\ref{e2a}), by definition, does not have upper limit. We can also calculate the second variation $\delta^2 A_M=\int_{t_a}^{t_b}\delta^2 p\dot{x}dt=2m\int_{t_a}^{t_b}\delta\dot{x}^2dt\geq 0$, which proves the minimum of the vanishing first variation $\delta A_M=0$.

As expected, this dissipative MPLA makes it possible to study easily the connection between PLA and the optimization of energy dissipation. Let us first show an interesting case where the motion is damped by Stokes drag $f_d=-m\zeta\dot{x}$.  The variation of the Maupertuis action can be written as
\begin{eqnarray}                                              \label{e7b}
\delta A_M=\delta\frac{1}{\zeta}\int_{x_a}^{x_b}m\zeta\dot{x}dx=\frac{\delta E_d^b}{\zeta}.
\end{eqnarray}
where $E_d^b=\int_{x_a}^{x_b}\zeta\dot{x}dx$ is the quantity of the work of the friction force over the entire trajectory from $a$ to $b$. Hence the MPLA $\delta A_M=0$ entails $\delta E_d^b=0$, i.e., least dissipation. In other words, {\it the path of least action is just the path of least resistance}.

The connection between MPLA and optimization of dissipation can be investigated in a more general way.
Let $f_d=-f(\dot{x})$ be a certain friction force where $f(\dot{x})$ is a positive increasing function of the magnitude of velocity. Let $E_d^b=-\int_{x_a}^{x_b}f_ddx=\int_{x_a}^{x_b}f(\dot{x})dx$ be the dissipated energy over the entire trajectory from $a$ to $b$. Changing the integral variable into time gives $E_d^b=\int_{0}^{T}f(\dot{x})\left|\dot{x}\right|dt=\int_{0}^{T}P(\dot{x})dt$ where $P(\dot{x})=f(\dot{x})\left|\dot{x}\right|$ is the power of the friction force to do work and should be always positive and increasing function of $\left|\dot{x}\right|$. We can write $P(\dot{x}^2)=P(y)$ with $y=\dot{x}^2$. The variations $\delta x$ and $\delta\dot{x}$ yield a variation of $E_b^a$ as follows

\begin{equation}                                            \label{e80}
\delta E_d^b=\int_{x_a}^{x_b}\frac{dP(y)}{dy}2\dot{x}\delta\dot{x}dt
=\int_{x_a}^{x_b}\frac{dP(y)}{dy}(\dot{x}\delta\dot{x}+\dot{x}\delta\dot{x})dt
\end{equation}
We write the first term in the parentheses as fllows $P'(y)\dot{x}\delta\dot{x}=P'(y)\delta(\frac{1}{2}\dot{x}^2)=-\frac{P'(y)}{m}(\frac{\partial V_1}{\partial x}+\frac{\partial E_d}{\partial x})\delta x$ where $P'(y)=\frac{dP(y)}{dy}$. We have used the conservation of total energy $\delta H=\delta(K_1+V_1+E_d)=0$. For the second term, we make a time integral by parts and use the same conditions as in the passage from Eq.(\ref{e2}) to Eq.(\ref{e3}) to write $\int_{0}^{T}P'(y)\dot{x}\delta\dot{x}dt =-\int_{0}^{T}\frac{d}{dt}[P'(y)\dot{x}]\delta xdt=-\int_{0}^{T}[\frac{dP'(y)}{dt}\dot{x}+P'(y)\ddot{x}]\delta xdt$. Finally, we have
\begin{equation}                                            \label{e8}
\delta E_d^b=-\int_0^T\left[\frac{dP'(y)}{dt}m\dot{x}+P'(y)(\frac{\partial V_1}{\partial x}+\frac{\partial E_d}{\partial x}+m\ddot{x})\right]\delta xdt.
\end{equation}
Since $m\ddot{x}=-\frac{\partial V_1}{\partial x}-\frac{\partial E_d}{\partial x}$ as a consequence of the MPLA $\delta A_M=0$, we see that $\delta E_d^b=-\int_0^T\frac{dP'(y)}{dt}m\dot{x}\delta xdt$ which is not vanishing, in general. So $E_d^b$ does not have stationarity on the least action path or, in other words, this latter is not necessarily the path of least dissipation or resistance. This is our general result.

However, it is clear from Eq.(\ref{e8}) that if $\frac{dP'(y)}{dt}=0$, $\delta E_d^b$ vanishes whenever $\delta A_M=0$ with $m\ddot{x}=-\frac{\partial V_1}{\partial x}-\frac{\partial E_d}{\partial x}$. There are many motions satisfying this condition. The first one is the Stokes' drag giving $P=m\zeta\dot{x}^2=m\zeta y$ and $\frac{dP'(y)}{dt}=\frac{d(m\zeta)}{dt}=0$. Since $P(y)$ is an increasing function of the magnitude of velocity or of $y$, $P'(y)=\frac{dP(y)}{dy}$ is always positive, so $\delta E_d^b=0$ is a minimum just as $\delta A_M=0$. This is in accordance with the conclusion obtained above with Eq.(\ref{e7b}).

The second motion leading to $\frac{dP'(y)}{dt}=0$ is the motion with constant velocity, be it along a straight or curved path and whatever is the velocity dependence of the power $P$. From Eq.(\ref{e8}), we still have the equivalence between MPLA and least dissipation, i.e., $\delta E_d^b=0$ following $\delta A_M=0$. This conclusion can also be reached from the following analysis. By definition, the action $A_M=\int_a^b pds$, ($dx$ is replaced by $ds$ for curved path where $s=f(t)$ is the moving equation). The result is $A_M=pL_{ab}$ where $L_{ab}=\int_a^b ds$ is the length of the path. Hence MPLA $\delta A_M=0$ entails $\delta L_{ab}=0$, i.e., the path of least distance. Since the velocity is constant, least distance implies least time $\delta T=0$ where $T=\int_a^b dt$ is the duration of the motion. On the other hand, as the friction force is constant, the energy dissipated between $a$ and $b$ is $E_d^b=f_d\int_a^bds=f_dL_{ab}$. The minimum distance $\delta L_{ab}=0$ then yields a minimum dissipation $\delta E_d^b=0$.

In summary, by considering a conservative system composed of the moving body and its environment coupled by friction, we have formulated a Maupertuis principle of least action for damped motion. This formalism allows to connect the optimization of energy dissipation to PLA. According to this formulation, the Maupertuis principle of least action is equivalent to the least dissipation or path of least resistance in the following two cases: 1) all motions damped by Stokes drag, and 2) all stationary motions with constant velocity. Otherwise, the paths of least action (solutions of the Newtonian equation) are not necessarily the least dissipative in energy. Since the Newtonian laws are universal for classical mechanical systems, this work means that the extremum rules such as least dissipation, least resistance, least effort, or maximum dissipation are not general laws for mechanical motions. However, the above two cases include a very large number of mechanical motions in Nature for which the Maupertuis action $A_M$ really represents the energy cost of motion. Mr. Maupertuis was partially right.

\end{document}